\documentclass[12pt]{article}
\usepackage{graphicx}
\begin{document}
\author{M. I. Jaghoub$^1$\thanks{
E-mail address: mjaghoub@ju.edu.jo} \hspace{0.3ex}, \
 M. F. Hassan$^1$, \  G. H. Rawitscher$^2$ \\
 {\small { $^1$ Department of Physics, University of Jordan, P. C. 11942,
Amman, Jordan}}\\
{\small {$^2$ Department of Physics, University of Connecticut, Storrs, CT
 06269, USA}}}
\title{Novel Source of Nonlocality in the Optical Model}
\maketitle

\begin{abstract}
In this work we fit $neutron$ - $^{12}C$ elastic scattering angular
distributions in the energy range $12$ to $20$ MeV, by adding a velocity
dependent term to the optical potential. This term introduces a wave
function gradient, whose coefficient is real and position dependent, and
which represents a nonlocality. We pay special attention to the prominent
backscattering minima which depend sensitively on the incident energies, and
which are a tell-tale of nonlocalities. Reasonable fits to the analyzing
power data are also obtained as a by-product. All our potentials have the form of conventional Woods - Saxon shapes or their derivatives. The number of our parameters (12) is smaller than the number for other local optical  potentials, and they vary monotically with energy, while the strengths of the real and imaginary parts of the central potential are nearly constants. Our nonlocality is in contrast to other forms of nonlocalities introduced previously.  

\end{abstract}

PACS numbers: 24.10.Ht, 25.40.Dn, 24.70.+s

\newpage

\section{Introduction}

Since its inception in 1954 \cite{Feschbach_1954} the nuclear optical model
has undergone many improvements and refinements. Two main theoretical
approaches emerged. One, the microscopic non relativistic approach, that
consists in folding a two-body nucleon-nucleon g-matrix into the nucleon
distributions contained in the target nucleus \cite{Kerman}, and iteratively
includes the many-body aspects of the nucleus \cite{Crespo}. The other
approach makes use of the Dirac equation to describe the wave function of
the incident nucleon, and is of a more phenomenological nature \cite{Clark}. Both
approaches provide almost equally good descriptions of the nucleon
scattering cross sections and polarizations in a range of energies, as
demonstrated recently for a representative number of target nuclei \cite{Deb}. The Dirac approach automatically introduces a nonlocality through the
Darwin term, but does not allow for the Pauli exclusion principle, while the
non relativistic approach does include the latter \cite{Deb}. It is also
found that the Dirac approach requires a much smaller energy dependence of
the Dirac optical potentials than the corresponding non relativistic
potentials, which suggests that this difference is related to the
nonlocality present in the Dirac procedure \cite{clark-1, RAWITSCHER}.

There should be at least two sources of nonlocality in the conventional, non relativistic
optical model: The first is due to the Pauli exclusion principle. This
nonlocality is usually taken into account by writing the overall wave
function as a Hartree-Fock determinant, and it leads to an additional
integral term in the Schr\"{o}dinger equation. Other methods also exist \cite{Kamos}.  The second is a "Feshbach
nonlocality", that is due to the coupling of the inelastic excitations to
the ground state in the elastic channel during the scattering process. Since
the form of this nonlocality is difficult to quantify, it usually is
ignored, or taken into account by numerically coupling a few inelastic
channels to the elastic channel \cite{Geramb}. Another method, proposed by
Perey and Buck \cite{Perey}, is to assume an "ad hoc" nonlocality expression
in terms of a combination of exponential functions of position, and obtain
the local equivalent potential in the Schr\"{o}dinger equation. The
Perey-Buck nonlocality is found to provide an energy dependence of the
central local equivalent optical potential that is in agreement with the
phenomenological energy dependence, and leads to Perey Damping factors that
are roughly similar to the ones in the Dirac optical model approach \cite
{RAWITSCHER}. Other forms of nonlocalities have also been explored, notable
among them being a parity dependent term in the optical model \cite
{Mackintosh, Cooper}, which led to a tentative justification in terms of the
Feshbach channel coupling effect \cite{RG_94, Rawitscher-L}. In a recent work, a nonlocal optical potential which has the Perey-Buck type spatial dependence  was employed in the calculations  of three-body direct  nuclear reactions. An important nonlocality effect has been found  for some transfer reactions \cite{Detluva}.

It is important to understand the presence of nonlocalities, because the
corresponding local representation of the optical model leads to a phase
equivalent wave function that is larger in magnitude than the corresponding
nonlocal wave function (the Perey damping factor), and hence leads to
discrepancies in the distorted wave approximation calculations (DWBA) of
inelastic or rearrangement processes, which in turn lead to errors in
obtaining the shell-model occupation numbers of the target nucleus.

The purpose of this paper is to introduce yet another nonlocality term into
the optical potential, that is expressed by the presence of a derivative
term in the Schr\"{o}dinger equation. The coefficient is a position-dependent
effective mass of the nucleon embedded in the nuclear medium, resulting from
the interaction with the other nucleons. The concept of an effective mass is
relevant to the nuclear many body problem in connection with the
energy-density functional approach. Here, the nonlocal terms are
usually interpreted as a three-dimensional position-dependent effective mass 
\cite{Ring}. By considering the most general kinetic energy operator for a
particle with a spatially variable mass \cite{Roos}, the Schr\"{o}dinger
equation can be recast in a form that describes a constant mass moving in a
velocity-dependent potential \cite{JaghoubA28}. Our preliminary
justifications for this nonlocality is based on the excellent fits to
low-energy $N-^{12}C$ elastic scattering data which we achieve. In
particular, the fits reproduce very well the pronounced minima corresponding
to large angle backward scattering, which are usually associated with
nonlocalities \cite{Clark}. Even though we did not include the analyzing
power in the search for parameters in the fit to the elastic cross sections,
we also obtained reasonably good fits to the experimental analyzing powers,
a result that supports the presence of our velocity dependent term. It must
be stressed, however, that our velocity dependent term is not similar \cite
{MJ-GR} to the velocity dependent Darwin term that occurs in the Dirac based
formulation \cite{Arnold}, because the Darwin term is complex, while our
effective mass is real. Further, the Darwin term is closely
coupled to the spin orbit potential in the Dirac formulation, while our spin
orbit term is completely independent of the velocity term \cite{MJ-GR}. In a
future investigation \cite{MJ-GR} we will study the physical origin of our
nonlocality, in particular, whether it approximately simulates the effect of
channel coupling to excited states, or whether it simulates exchange effects, or introduces a new type of nonlocality. 

Velocity-dependent potentials have long been used in nuclear and atomic
physics. For example, such a potential was introduced to explain the
predominantly $p$-wave nature of the pion-nucleon scattering \cite
{Kisslinger}. In addition, a model assuming a velocity-dependent
nucleon-nucleon interaction reproduced the $^{1}S$, $^{1}D$ and $^{1}G$
singlet-even phase shifts required for the description of nucleon-nucleon
scattering cross sections \cite{Levinger}. In the field of atomic physics
the scattering of electrons from atomic oxygen and neon was studied in the
frame work of an analytic velocity-dependent potential \cite{Green}. In
addition, the effective mass formulation has been introduced into condensed
matter physics to describe the dynamics of electrons in semiconductor
hetrostructures such as compositionally graded crystals \cite{Geller} and
quantum dots \cite{Serra}. Further, scattering of electrons on disordered
double-barrier hetrostructures has been considered in Ref. \cite{Gomez}.
Finally, one of us presented  perturbation formalisms that accounts for the
effect of a small perturbing velocity-dependent potential on the bound-state
energies and the scattering phase shifts. \cite{Jaghoub27,JaghoubPR}.

\section{The velocity-dependent term \label{sec:VDP}}

In this section we shall briefly outline the effective mass formalism that
leads to a velocity-dependent term. As mentioned above, the most general
kinetic energy term used to describe a spatially variable mass $m(r)$ is
given as \cite{Roos} 
\begin{equation}
T=-\frac{\hbar ^{2}}{2m}\left[ m^{\delta }(r)\, \nabla \,m^{\beta }(r)\,
\nabla \,m^{\gamma }(r)+m^{\gamma }(r)\, \nabla \,m^{\beta }(r)\, \nabla
\,m^{\delta }(r)\right] .
\end{equation}
Since the mass depends on position it no longer commutes with the momentum
operator and the ambiguity parameters obey the constraint $\delta +\beta
+\gamma =-1$ \cite{Roos}. Interesting works were carried out to determine a
unique set of ambiguity parameters. For example, by considering the
one-dimensional Schr\"{o}dinger equation for a spatially variable mass $m(x)$
it was suggested that there is a privileged ordering, namely, $\delta
=0,\beta =-1$ which was obtained by demanding $[m(x)]^{-1}\partial /\partial
x$ be continuous at the point of discontinuity of $m(x)$ \cite{Leblond}. For
this set of parameters, the potential functions for the one- and
three-dimensional Schr\"{o}dinger equations were obtained in addition to
explicit expressions for the bound state energy spectrum and the
corresponding wave functions \cite{Haidari}. In addition, the formalism of
supersymmetric quantum mechanics is extended to the one-dimensional Schr\"{o}dinger equation with a spatially variable mass \cite{Plastino}.

Using the above form of $T$ and choosing the same set of parameters ($\delta
=0,\beta =-1$) the corresponding radial Schr\"{o}dinger equation reads: 
\begin{equation}
-\frac{\hbar ^{2}}{2m}\left \{ \frac{d^{2}}{dr^{2}}-\frac{m^{\prime }}{m}
\left[ \frac{d}{dr}-\frac{1}{r}\right] -\frac{l(l+1)}{r^{2}}\right \} v(k,r)=
\left[ E-V(r)\right] v(k,r),  \label{GenT}
\end{equation}
where $m\equiv m(r)$ and $v(k,r)=rR(r)$ is the reduced wave function and the
prime denotes a derivative with respect to $r$. By making the substitution 
\begin{equation}
\frac{1}{m}=\frac{1-\rho (r)}{m_{0}},  \label{effmass}
\end{equation}
where $\rho (r)$ is some isotropic function of the radial variable $r$ and $
m_{0}$ is a constant mass, equation (\ref{GenT}) reduces to 
\begin{equation}
(1-\rho )v^{\prime \prime }(r)-\left[ v^{\prime }(r)-\frac{v(r)}{r}\right]
\rho ^{\prime }-(1-\rho )\frac{l(l+1)}{r^{2}}v(r)=\frac{2m_{0}}{\hbar ^{2}}
\left[ V(r)-E\right] v(r),  \label{vdse}
\end{equation}
where the dependencies of the reduced wave function on $k$ and $\rho (r)$ on 
$r$ have been suppressed for clarity of presentation. This is exactly the
same equation one obtains when starting from the usual Schr\"{o}dinger
equation but with a velocity-dependent potential of the form: 
\begin{eqnarray}
\hat{V}(r,p) &=&V(r)+\frac{\hbar ^{2}}{2m_0}\mathbf{\nabla }\cdot \rho (r)
\mathbf{\nabla }  \nonumber \\
&=&V(r)+\frac{\hbar ^{2}}{2m_0}\left[ \rho (r)\nabla ^{2}+\nabla \rho
(r)\cdot \nabla \right]  \label{vdp}
\end{eqnarray}
The second term on the right hand side results in a kinetic energy term that
combines with the corresponding term in the Schr\"{o}dinger equation. The
third term, however, is proportional to the gradient of $\rho (r)$ in
addition to the gradient of the wave function. Once more, it is worth
mentioning that the gradient terms are not identical to those obtained when
the Dirac formalism is extended to the non relativistic regime.

\section{Velocity-dependent optical potential}

One of the nonlocalities that we propose to be present in the optical model
is due to the change in mass of the nucleon arising from its interactions
with other nucleons inside the nucleus. Our aim in this work is to determine
to what extent the inclusion of an "ad hoc" velocity-dependent term can
simulate such a nonlocality. For this purpose we shall introduce a gradient
part resulting in a velocity-dependent optical potential of the form given
in equation (\ref{vdp}) with the conventional optical model part given by, 
\begin{equation}
V(r)=-V_{0}f(r,x_{0})+4 i a_{w}W\frac{df(r,x_{w})}{dr}+20\left( \frac{\hbar }{
\mu c}\right) ^{2}(V_{so}+ i W_{so})\frac{1}{r}\frac{df(r,x_{so})}{dr}\vec{
\sigma}\cdot \vec{I},  \label{Vr}
\end{equation}
where $\mu$ is a constant neutron-nucleus reduced mass. For the gradient term we define 
\begin{equation}
\rho (r)=\rho _{0} \, a_{\rho }\frac{df(r,x_{\rho })}{dr},  \label{rhor}
\end{equation}
where $x_{0}$ stands for $(r_{0},a_{0})$ and so on for the rest of the
terms. The function $f(r,r_{j},a_{j})$ is of a Woods-Saxon form, namely, 
\begin{equation}
f(r,r_{j},a_{j})=\frac{1}{1+\exp [(r-r_{j}A^{1/3})/a_{j}]},
\end{equation}
where $A$ is the mass number of the target nucleus. Clearly, $\rho (r)$ is a
surface term, which may be interpreted as the gradient of the mass density
of the target nucleus \cite{Kisslinger}. In view of this, one would expect
the effect of including this term to be important closer to the nuclear
surface rather than deep in the interior of the target nucleus. In fact our
best fits are obtained with the peak of $\rho (r)$ close to the nuclear surface.

In an attempt to reduce the number of fit parameters we have used the same geometry parameters for the real and imaginary parts of the spin-orbit term. 
Our model has 12 fit parameters compared to the conventional optical potentials which have a number of fit parameters ranging from nine to twenty. For example, in a previous work \cite{Dave}, the conventional optical potential with nine fit parameters was used to fit the neutron-nucleus elastic angular distribution data for 1-$p$ shell nuclei ranging from $^{6}Li$ to $^{13}C$. The
bombarding neutron energies fell in the range  7 - 15 MeV, which does not include the pronounced large-angle backward scattering minima at 18 and 20 MeV. Further, as the authors stated, the emphasis was put on reproducing the overall behavior of the angular distributions and not the exact details as the low energy region is endowed with resonances. A more recent work
presented new global and local optical potentials for neutrons and protons
with incident energies ranging from 1 keV to 200 MeV for nuclei in the mass
range $24\leq A\leq 209$ \cite{Koning}. Each of the local optical potentials
included 20 fit parameters and resulted in excellent fits to
the experimental data. The strength of the central real potential showed the
largest variation as a function of the incident energy. Furthermore,
energy-dependent global Dirac optical potentials also resulted in excellent
fits to proton elastic scattering data corresponding to incident energies in
the range 20 - 1040 MeV for a number of light and heavy nuclei. The number
of optical potential parameters is about 24 and the data sets were
restricted to observables at angles less than 90 degrees so as to avoid the
effect of nonlocalities, which were believed to be important at large back
scattering angles \cite{Clark,Rawitscher-L}.

\section{Velocity-dependent optical potential fits}

As mentioned in the introduction there are sources of nonlocalities in the
non relativistic optical potential, such as exchange processes, and coupling
of the inelastic excitations to the ground state of the elastic channel. In
this work we introduce yet another source of nonlocality presumably
resulting from the change in mass of the incident nucleon due to its
interactions with the other nucleons inside the nucleus, and we ignore the
other two. In order to simulate such a "medium" nonlocality, we have added a
velocity-dependent term to the optical potential. This introduces a wave
function gradient term whose coefficient is real and position dependent. As
outlined in section~\ref{sec:VDP}, the Schr\"{o}dinger equation describing a
spatially variable mass can be made identical to a Schr\"{o}dinger equation
describing the dynamics of a constant mass moving in a velocity-dependent
potential of the form given in equation (\ref{vdp}). Our aim is to test the
ability of such a velocity-dependent optical potential to simulate the
effective mass nonlocality especially in the backward (large angle)
scattering region which has long been associated with nonlocalities 
\cite{Clark,Rawitscher-L}.

\begin{figure}[ht]
\includegraphics[height=60mm, width=80mm]{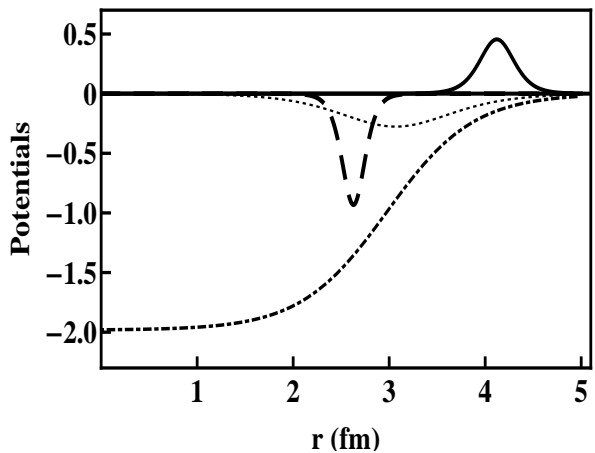}
\caption{The potential parts of the proposed velocity-dependent optical model at a neutron incident energy of 12 MeV. Central real part (dash - dotted),  central imaginary part (dotted), real part of the spin-orbit term (dashed) and, $\rho(r)$, the velocity-dependent term  (solid).   \label{potentials} }
\end{figure}

Using the proposed velocity dependent optical potential (VDOP) with $V(r)$
and $\rho (r)$ given by equations (\ref{Vr}) and (\ref{rhor}) respectively,
we searched for sets of parameters that fit the $N-^{12}C$ angular
distribution elastic scattering data in the low-energy $12-18$ MeV range.
Lower energies were not considered in order to avoid resonance effects. The
source of the data is the Evaluated Nuclear Data File \cite{ENDF}. The data
showed pronounced large angle backward scattering minima that are very
sensitive to energy changes, thus signifying sources of nonlocalities. We
varied the parameters until a best fit to the elastic angular distributions
are obtained. The fits are shown in Figures \ref{AngularFits1} and \ref{AngularFits2} and our best fit parameters
are presented in Tables~\ref{Table1} and \ref{Table2}. Clearly, the VDOP has
reproduced well the experimental data, most notably the pronounced large-
angle, backscattering minima at 18 and 20 MeV. The quality of the fits at
the large back angle minima is found to be sensitive to both the spin-orbit
and to the velocity-dependent terms. Further, the peak of the spin-orbit
term monotically moved into the nuclear interior as the incident energy
increased. As shown in Figure \ref{potentials}, at $12$ MeV incident energy, the peak of $\rho (r)$ is at the
outer edge of the real central potential.  However, as the energy
increased the peak moved into the nuclear interior but remained within the
region of the diffused edge of the central potential. This supports the
hypothesis that $\rho (r)$ is related to the gradient of the target's mass
density \cite{Kisslinger}. The strengths of the real and imaginary
potentials $V_{0}$ and $W$ were nearly stable as a function of incident
energy especially for $E\geq 16$ MeV. 

For a perfect theory, one would expect the fit parameters to have a smooth variation with energy. By inspecting Tables \ref{Table1} and \ref{Table2} it can be seen that our parameters generally vary smoothly with energy. However, some of the parameters at $E=16$ MeV are quite different form the corresponding ones at neighboring higher and lower energies. This may be due to the fact that there are several energy-dependent additional effects in the nucleon-nucleus elastic scattering process that are still explicitly left out at the present stage. One is the exchange effect, and the other is a channel coupling effect. The latter involves a Feschbach-like polarization potential, that is known to be angular momentum and energy dependent and is nonlocal. In a future work, it is our hope to take such effects into account by re-expressing both effects in terms of gradient terms, thus modifying our proposed velocity-dependent optical potential. This is expected to result in an even smoother variation of our parameters with energy.

Although we did not include the polarization data in the fitting procedure,  
our model makes reasonable theoretical predictions for the analyzing power $A_{y}(\theta)$ as can be seen in Figure \ref{ApFits}. At the energies of $11.9$ and $13.9$ MeV the experimental results are
given in reference \cite{Woye} and at$14.2$ MeV they are given by \cite
{Casparis}. Since the evaluated nuclear data file did not have angular
distributions data of $A_{y}(\theta )$ at the aforementioned elastic
scattering energies, we calculated the predicted polarization asymmetry
corresponding to $11.9$ MeV using our fit parameter values for the elastic
angular distribution obtained  at $12$ MeV. For the remaining $13.9$ and $
14.2$ MeV energies, we used our parameter values obtained for the fit to the
elastic angular distribution at $14.0$ MeV.  We expect
that our theoretical predictions for the analyzing power can be improved by including other sources of nonlocalities,
such as exchange effects and coupling of the ground state elastic channel to
inelastic excitations.

\begin{figure}[ht]
\includegraphics[]{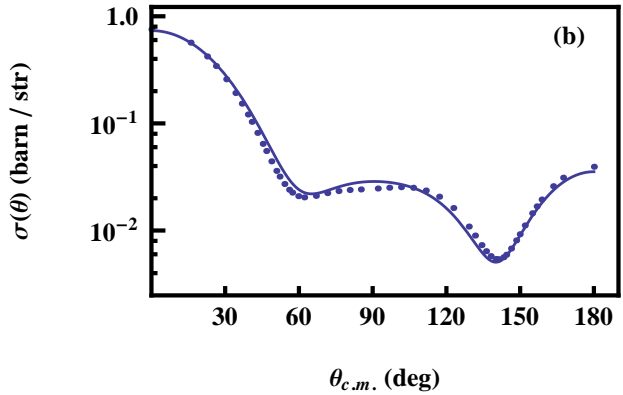}
\caption{The conventional optical model ($\rho(r)=0$) fit to the data at 14 MeV incident neutron energy.  \label{fit_rho=0} }
\end{figure}

\begin{figure}[ht]
\includegraphics[]{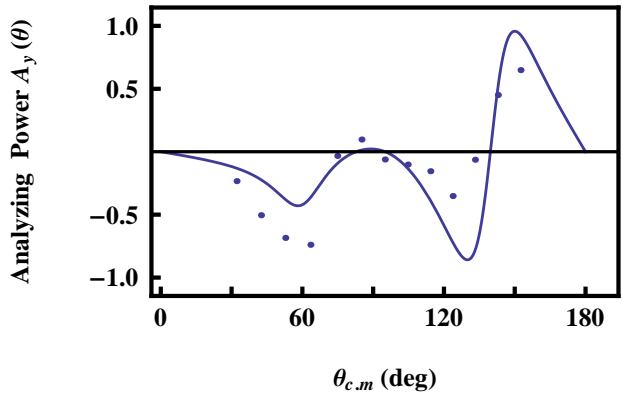}
\caption{The conventional optical model ($\rho(r)=0$) prediction for the analyzing power at 13.9 MeV.  \label{Apfit_rho=0} }
\end{figure}

We have also fitted the 14 MeV angular distribution elastic scattering data using a conventional optical potential model by setting the gradient term $\rho(r)$ to zero. This energy was chosen as analyzing power data are available at 13.9 and 14.2 MeV, which provide more physical constrains on the values of the potential parameters. The corresponding fit to the elastic angular distribution  is shown in Figure \ref{fit_rho=0}. Clearly, the overall behavior of the differential  cross section is reproduced but the fine details are less well described compared to the VDOP fit at the same energy as shown in Figure \ref{AngularFits1} (b). In addition, fits to the data became harder to achieve as the incident energy increased. In particular, the large-angle backward scattering minima  at 18 and 20 MeV and the details of the differential cross sections were much less well reproduced compared to the corresponding VDOP fits. Further, the theoretical prediction for the analyzing power, $A_y(\theta)$, presented in Figure ~\ref{Apfit_rho=0} still needs to be improved, as was the case for the VDOP, but  agreement with experimental data around $60^0$ is clearly better for the prediction of the proposed velocity-dependent optical potential.

\section{Discussion and conclusion}

\begin{figure}[ht]
\includegraphics[height=120mm, width=90mm]{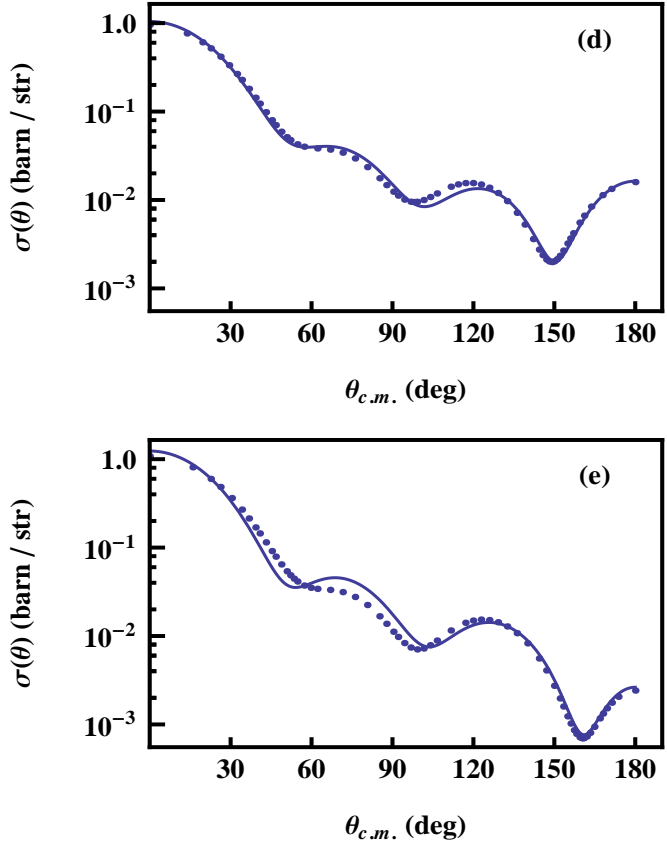}
\caption{The velocity-dependent optical potential fits for $N-^{12}C$ elastic scattering at 18 (d) and 20 (e) in units of MeV. The pronounced minima at large backward scattering angles are evident. The model parameters are given in Tables \ref{Table1} and \ref{Table2}. The data is obtained from reference \cite{ENDF}. \label{AngularFits2} }
\end{figure}

In a previous work Cooper and Mackintosh \cite{Mackintosh} showed that good fits
to low energy scattering of either protons or neutrons from $^{12}C$ or $
^{16}O$ out to $180^{0}$ could not be accomplished using only local optical
potentials. They introduced a nonlocality by means of parity-dependent
potentials which contained factors $(-)^{L}$, where $L$ is the orbital
angular momentum of the projectile relative to the target nucleus. In this
work we introduce a different type of nonlocality in the form of a derivative term. This nonlocality, described by a position-dependent
effective mass which multiplies the first order derivative term of the
scattering wave function, is presumed to be a consequence of the static
interaction of the incident nucleon with the other nucleons in the nuclear
medium. Its physical origin will be the object of a separate investigation 
\cite{MJ-GR}. One argument justifying the introduction of our velocity
dependent term, is that the magnitude of the central part of the optical
potential depends much less on energy  than for fits using only local
potentials. Another argument is that we obtain excellent fits to the
pronounced minima, especially at 18 and 20 MeV, in the angular distribution of the elastic cross section
at large angles, shown in Figure \ref{AngularFits2}, that can otherwise not be fitted
well \cite{Mackintosh,Rawitscher-L}. Furthermore, our predicted
analyzing powers have trends that are in agreement with experiment.

\begin{figure}[ht]
\includegraphics[height=160mm, width=90mm]{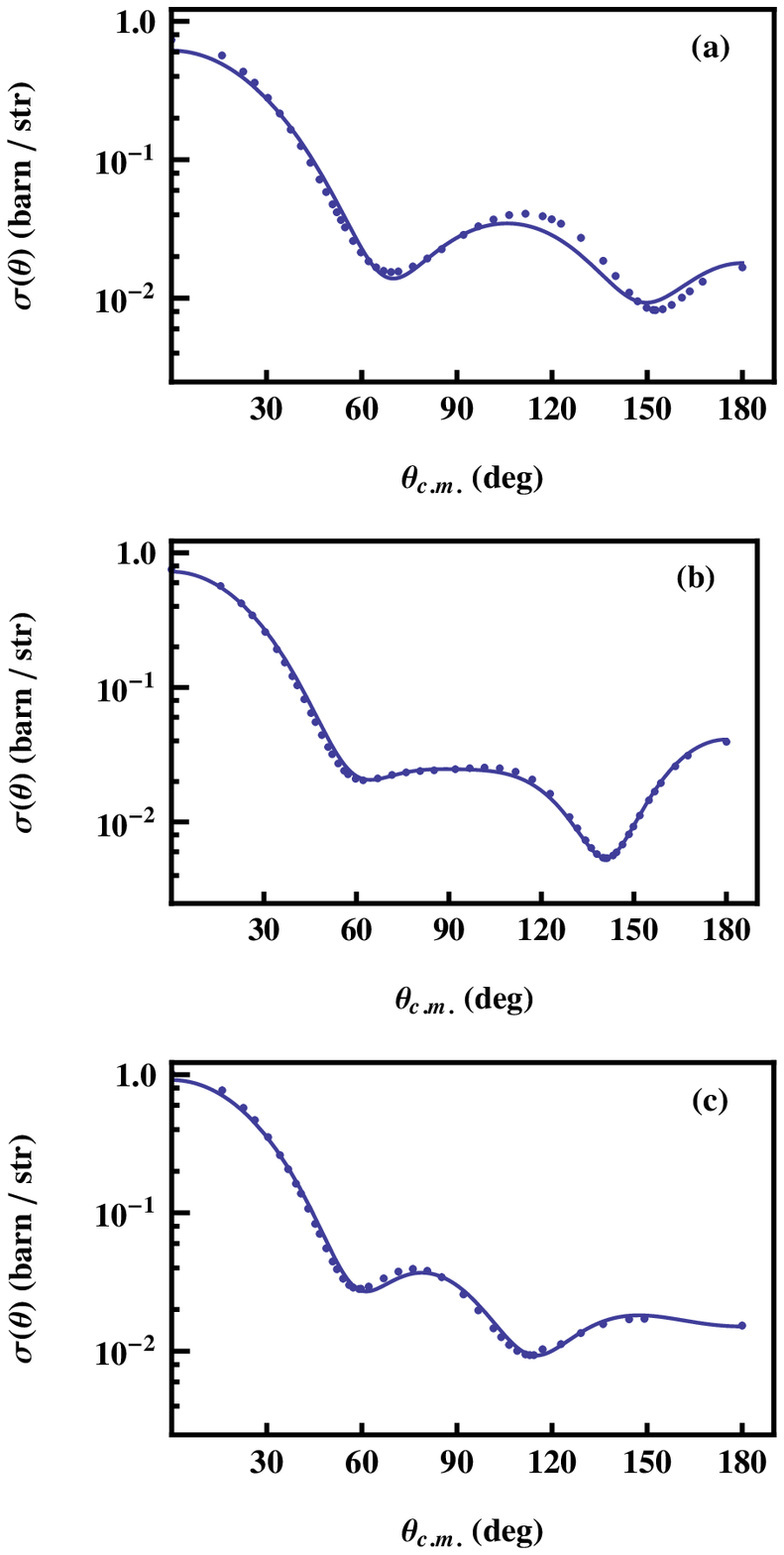}
\caption{The velocity-dependent optical potential fits for $N-^{12}C$ elastic scattering at 12 (a), 14 (b) and 16 (c) all in units of MeV. The model parameters are given in Tables \ref{Table1} and \ref{Table2}. The data is obtained from reference \cite{ENDF}. \label{AngularFits1} }
\end{figure}

\begin{figure}[ht]
\includegraphics[height=150mm, width=80mm]{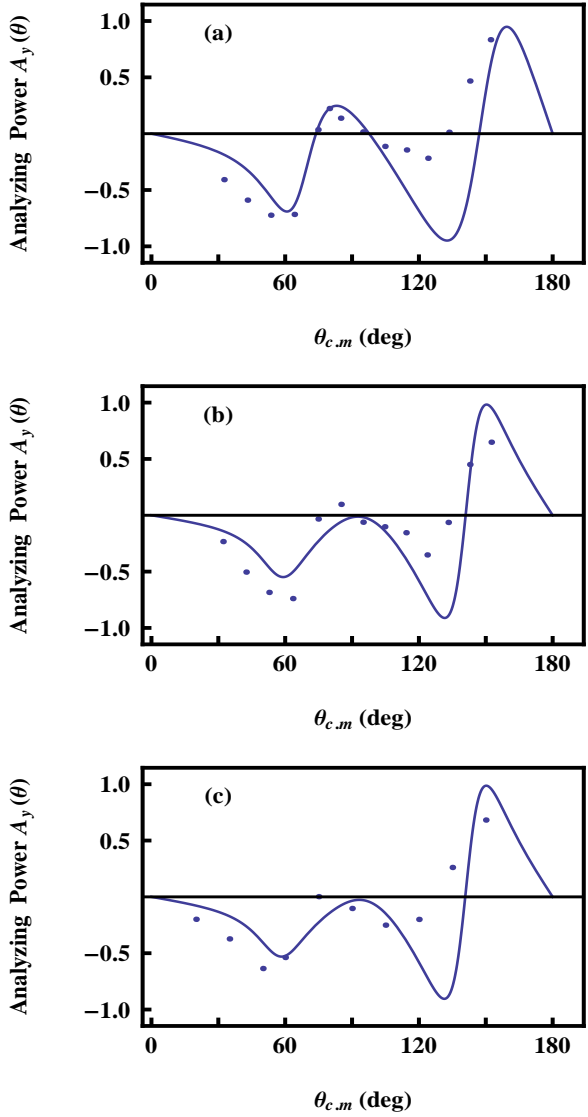}
\caption{The velocity-dependent optical potential predictions for the analyzing power at different incident neutron energies. The experimental data for 11.9 (a) and 13.9 (b) MeV energies are taken from Ref. \cite{Woye} while for the 14.2 MeV (c) the data is taken from Ref. \cite{Casparis}. \label{ApFits} }
\end{figure}

The proposed velocity-dependent potential has a total number of 12 fit parameters given in Tables~\ref{Table1} and \ref{Table2}. This is compared to conventional optical models where the typical number of fit parameters ranges between nine and twenty \cite{Dave,Koning}.  In this work, the details of the differential cross section fits were sensitive to both
the spin-orbit and velocity-dependent terms, in particular to their
strengths and radii. The best fits, shown in Figures \ref{AngularFits1} and \ref{AngularFits2}, were obtained with the peak of $\rho (r)$
located within the region of the diffused nuclear surface. This supports the
hypothesis that $\rho (r)$ may be viewed as the gradient of the nuclear mass
density \cite{Kisslinger}. By inspecting Tables~\ref{Table1} and \ref{Table2}
, it is clear that the radius of the spin-orbit term decreased monotonically
as the incident energy increased. In addition, the fit values for the
strengths of the real and imaginary parts $V_{0}$ and $W$ showed only a
small energy dependence. The theoretical predictions of our model for the analyzing power given in Figure \ref{ApFits} can probably be improved by including other sources of nonlocalities such as exchange processes and channel coupling to inelastic excitations.

\begin{table}[ht]
\begin{tabular}{|l|c|c|c|c|c|c|}
\hline
$E_{lab}$ & $U_{0}$ & $r_{0}$ & $a_{0}$ & $W$ & $r_{w}$ & $a_{w}$ \\ 
$MeV$ & $MeV$ & $fm$ & $fm$ & $Mev$ & $fm$ & $fm$ \\ \hline \hline
$12$ & $44.5$ & $1.30$ & $0.45$ & $6.2$ & $1.34$ & $0.39$ \\ \hline
$14$ & $44.7$ & $1.31$ & $0.46$ & $5.7$ & $1.30$ & $0.45$ \\ \hline
$16$ & $40.1$ & $1.39$ & $0.37$ & $6.3$ & $1.47$ & $0.23$ \\ \hline
$18$ & $40.1$ & $1.31$ & $0.56$ & $6.6$ & $1.32$ & $0.40$ \\ \hline
$20$ & $39$ & $1.35$ & $0.51$ & $6.5$ & $1.36$ & $0.47$ \\ \hline
\end{tabular}
\caption{Velocity-dependent optical model fit parameters for the $N-^{12}C$
angular distribution elastic scattering. The potential terms are given by
equations (\protect \ref{Vr}) and (\protect \ref{rhor})}
\label{Table1}
\end{table}

\begin{table}[ht]
\begin{tabular}{|l|c|c|c|c|c|c|}
\hline
$E_{lab}$ & $V_{so}$ & $r_{so}$ & $a_{so}$ & $\rho _{0}$ & $r_{\rho }$ & $
a_{\rho }$ \\ 
$MeV$ & $MeV$ & $fm$ & $fm$ & $-$ & $fm$ & $fm$ \\ \hline \hline
$12$ & $16.99+2.80\ i$ & $1.15$ & $0.08$ & $-1.82$ & $1.80$ & $0.12$ \\ 
\hline
$14$ & $11.59+3.67\ i$ & $1.03$ & $0.11$ & $-1.71$ & $1.50$ & $0.13$ \\ 
\hline
$16$ & $24.72+4.83\ i$ & $1.00$ & $0.11$ & $-1.46$ & $1.30$ & $0.15$ \\ 
\hline
$18$ & $23.56+15.45\ i$ & $0.81$ & $0.15$ & $-2.27$ & $1.33$ & $0.15$ \\ 
\hline
$20$ & $17.19+21.25\ i$ & $0.80$ & $0.19$ & $-2.05$ & $1.20$ & $0.15$ \\ 
\hline
\end{tabular}
\caption{Velocity-dependent optical model fit parameters for the $N-^{12}C$
angular distribution elastic scattering. The potential terms are given by
equations (\ref{Vr}) and (\ref{rhor})}
\label{Table2}
\end{table}

In summary, we have introduced into the optical model potential a
nonlocality that is normally not used to describe nucleon-nucleus elastic
scattering. It is in the form of a derivative term, multiplied by a function
that depends on a position dependent effective mass. Our excellent fits to
the elastic angular distribution, especially at large angles, justify
engaging in an exploration of how our nonlocality is related to the
conventional nonlocalities due to exchange effects and coupling to
inelastic channels, or whether our nonlocality is indeed related to a change
of effective mass of the projectile in the medium of the target nucleus, and
hence represents a new effect.  

\end{document}